\documentclass[a4paper,11pt]{article}
\usepackage{pos}
\usepackage{subfigure} 
\usepackage[skip=0pt]{caption}

\title{Double beta decay results from the CUPID-0 experiment}

 \author*[a,b]{D.~Chiesa}
 \author[c]{O.~Azzolini}
 \author[d]{J.~W.~Beeman}
 \author[e,f]{F.~Bellini}
 \author[a,b]{M.~Beretta}
 \author[b]{M.~Biassoni}
 \author[a,b]{C.~Brofferio}
 \author[g]{C.~Bucci}
 \author[a,b]{S.~Capelli}
 \author[f]{L.~Cardani}
 \author[g,h]{E.~Celi}
 \author[a,b]{P.~Carniti}
 \author[f]{N.~Casali}
 \author[a,b]{M.~Clemenza}
 \author[b]{O.~Cremonesi}
 \author[f]{A.~Cruciani}
 \author[g,h]{A.~D'Addabbo}
 \author[f]{I.~Dafinei}
 \author[i,j]{S.~Di~Domizio}
 \author[f,h]{F.~Ferroni}
 \author[a,b]{L.~Gironi}
 \author[k]{A.~Giuliani}
 \author[g]{P.~Gorla}
 \author[a,b]{C.~Gotti}
 \author[c]{G.~Keppel}
 \author[e,f]{M.~Martinez} 
 \author[g,h]{S.~Nagorny} 
 \author[a,b]{M.~Nastasi}
 \author[g]{S.~Nisi}
 \author[l]{C.~Nones}
 \author[g]{D.~Orlandi}
 \author[a,b]{L.~Pagnanini}
 \author[i,j]{M.~Pallavicini}
 \author[g,h]{L.~Pattavina} 
 \author[a,b]{M.~Pavan}
 \author[b]{G.~Pessina}
 \author[e,f]{V.~Pettinacci}
 \author[g]{S.~Pirro}
 \author[a,b]{S.~Pozzi}
 \author[a,b]{E.~Previtali}
 \author[g,h]{A.~Puiu}
 \author[g,m]{C.~Rusconi} 
 \author[g,h]{K.~Sch\"affner}
 \author[f]{C.~Tomei}
 \author[e,f]{M.~Vignati}
 \author[k]{A.~Zolotarova} 

\affiliation[a]{Dipartimento di Fisica, Universit\`{a} di Milano - Bicocca, Milano I-20126 - Italy}
\affiliation[b]{INFN - Sezione di Milano - Bicocca, Milano I-20126 - Italy}
\affiliation[c]{INFN - Laboratori Nazionali di Legnaro, Legnaro (Padova) I-35020 - Italy}
\affiliation[d]{Materials Science Division, Lawrence Berkeley National Laboratory, Berkeley, CA 94720 - USA}
\affiliation[e]{Dipartimento di Fisica, Sapienza Universit\`{a} di Roma, Roma I-00185 - Italy}
\affiliation[f]{INFN - Sezione di Roma, Roma I-00185 - Italy}

\affiliation[g]{INFN - Laboratori Nazionali del Gran Sasso, Assergi (L'Aquila) I-67010 - Italy}
\affiliation[h]{Gran Sasso Science Institute, 67100, L'Aquila - Italy}
\affiliation[i]{Dipartimento di Fisica, Universit\`{a} di Genova, Genova I-16146 - Italy}
\affiliation[j]{INFN - Sezione di Genova, Genova I-16146 - Italy}
\affiliation[k]{Universit\'e Paris-Saclay, CNRS/IN2P3, IJCLab, 91405 Orsay, France}
\affiliation[l]{IRFU, CEA, Universit\'{e} Paris-Saclay, F-91191 Gif-sur-Yvette, France}
\affiliation[m]{Department of Physics and Astronomy, University of South Carolina, Columbia, SC 29208 - USA}

\emailAdd{davide.chiesa@mib.infn.it}

\abstract{A convincing observation of neutrino-less double beta decay (0$\nu$DBD) relies on the possibility of operating high energy-resolution detectors in background-free conditions.
Scintillating cryogenic calorimeters are one of the most promising tools to fulfill the requirements for a next-generation experiment. Several steps have been taken to demonstrate the maturity of this technique, starting from the successful experience of CUPID-0.
The CUPID-0 experiment demonstrated the complete rejection of the dominant alpha background measuring the lowest counting rate in the region of interest for this technique. Furthermore, the most stringent limit on the $^{82}$Se 0$\nu$DBD was established running 26 ZnSe crystals during two years of continuous detector operation.
In this contribution we present the final results of CUPID-0 Phase I including a detailed model of the background, the measurement of the $^{82}$Se 2$\nu$DBD half-life and the evidence that this nuclear transition is single state dominated. 
}

\FullConference{%
  40th International Conference on High Energy physics - ICHEP2020\\
  July 28 - August 6, 2020\\
  Prague, Czech Republic (virtual meeting)
}

\begin{document}
\maketitle

\section{Introduction}

Neutrinoless double beta decay (0$\nu$DBD) is a hypothesized spontaneous nuclear transition in which a (A,Z) nucleus decays to a more stable (A,Z+2) one emitting 2 electrons and 0 neutrinos. Unlike the two-neutrino double beta decay (2$\nu$DBD), which is a second order transition allowed in the framework of the Standard Model (SM) of particle physics, 0$\nu$DBD has never been measured. The observation of this decay would imply that lepton number is not conserved and that neutrinos are massive Majorana particles~\cite{Majorana}, pointing to new physics beyond the SM. 
This is why there is a big interest in the search for this extremely rare decay. 
Today, the most sensitive experiments are setting lower limits on the 0$\nu$DBD half-lives ($T_{1/2}$) of about 10$^{25}$ $-$ 10$^{26}$ yr~\cite{annurev-nucl-101918-023407}.
In 0$\nu$DBD, the sum of the kinetic energies of the two emitted electrons and of the recoiling nucleus is equal to the Q-value of the transition (Q$_{\beta\beta}$).
Therefore, the signature of 0$\nu$DBD is a peak at Q$_{\beta\beta}$ in the kinetic energy spectrum. 
Next generation experiments aim at increasing the sensitivity up to $T_{1/2}$ of the order of 10$^{27}$~yr. 
To reach this goal, the detectors must be capable of monitoring $\mathcal{O}(10^{27})$ double beta decaying nuclei for a few years, and must feature high energy resolution and background close to zero at the ton$\times$yr exposure scale around Q$_{\beta\beta}$.

Cryogenic calorimeters are one of the most promising technology to fulfill these requirements~\cite{Artusa:2014wnl}. They consist of dielectric and diamagnetic crystals cooled at cryogenic temperatures ($\sim$10 mK) and equipped with thermistors to measure the temperature rise that occurs when a particle interaction deposits energy in the crystal volume. 
The crystals suitable for 0$\nu$DBD searches are those containing double beta decaying isotopes. Since the source is embedded in the detector, a very high detection efficiency is achieved. Moreover these detectors feature an excellent energy resolution $\Delta E / E \sim 10^{-3}$, and can be scaled in mass up to 1 tonne, as proved by the CUORE experiment, which is now taking data at the Laboratori Nazionali del Gran Sasso (LNGS, Italy) to search for 0$\nu$DBD of $^{130}$Te~\cite{CUORE:2019jhp}.

CUPID (CUORE Upgrade with Particle IDentification) is a proposed 0$\nu$DBD next generation experiment that will profit from the CUORE experience and cryogenic infrastructure to operate an array of 1500 Li$_2^{100}$MoO$_4$ scintillating calorimeters enriched to >95\% in $^{100}$Mo~\cite{CUPID:2019inu}. 
Together with the isotopic enrichment in the double beta decaying isotope, the other major upgrade with respect to CUORE will be the particle identification capability based on the simultaneous readout of the heat and light signal produced in scintillating crystals. 
This feature will be exploited to suppress the background produced by radioactive contaminations of the experimental setup that emit $\alpha$ particles degraded in energy, representing the dominant contribution to the background in CUORE~\cite{Alduino:2017qet}.
The first demonstrator experiments of the new technologies that will be implemented in CUPID are CUPID-0, which started data taking in 2017 at LNGS, and CUPID-Mo~\cite{CUPID-Mo:2019loe}, running at the Modane Laboratory (France) since 2019.

\section{The CUPID-0 experiment}

\begin{figure}[t!]
\begin{center}
\subfigure{\includegraphics[height=0.24\textwidth]{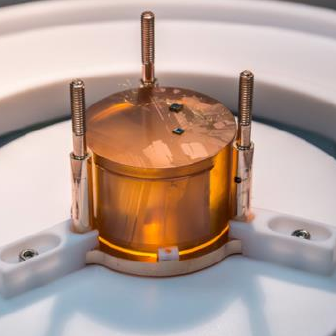}}
\subfigure{\includegraphics[height=0.24\textwidth]{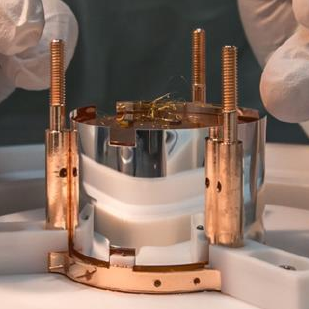}}
\subfigure{\includegraphics[height=0.24\textwidth]{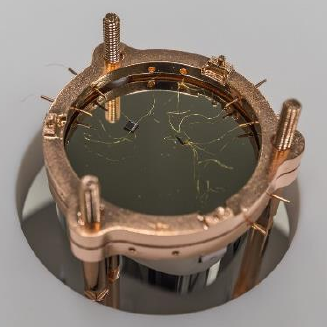}}
\subfigure{\includegraphics[height=0.24\textwidth]{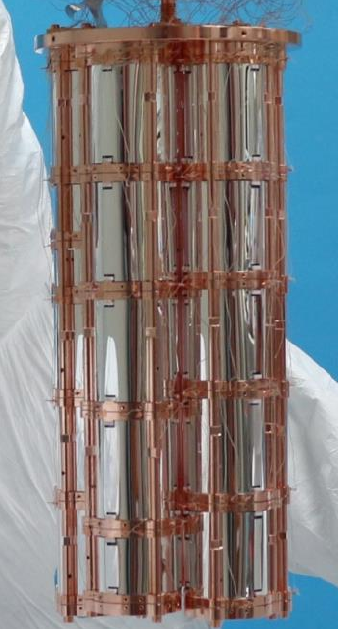}}
\end{center}
\caption{Pictures of the CUPID-0 detector. From left to right: a ZnSe crystal, the same crystal surrounded by the reflecting foil, the Ge light detector mounted on top, the CUPID-0 array of 26 scintillating calorimeters. }
\label{Fig:Detector} 
\end{figure}

CUPID-0, besides being a demonstrator for CUPID, is also a competitive  0$\nu$DBD experiment for the study of the double decay of $^{82}$Se.
The Q$_{\beta\beta}$ of $^{82}$Se (2997.9$\pm$0.3~keV), like the $^{100}$Mo one, lies in the energy range above the 2615~keV line of $^{208}$Tl where the $\beta/\gamma$ background from natural radioactivity is significantly lower.

The CUPID-0 detector has been installed at LNGS in the same cryostat previously used for the CUORE-0 experiment~\cite{CUORE-0:2015wka}. 
The detector is an array of 26 ZnSe scintillating calorimeters with a total mass of 10.5~kg, comprising 5.17~kg of $^{82}$Se thanks to the isotopic enrichment to $\sim$95\% of 24 crystals. 
To detect the light signal, each ZnSe crystal is equipped with two Ge crystals (4.4~cm in diameter and 170~$\mu$m thick) placed in correspondence with the top and bottom crystal faces. 
The Ge light detectors are operated as cryogenic calorimeters and, thanks to their small thermal capacitance, they are suitable for measuring low-energy light pulses. 
Both ZnSe and Ge crystals are equipped with neutron transmutation doped (NTD) temperature sensors to convert the heat signal into a measurable change in voltage proportional to the particle energy deposit. All crystals are held in position
by means of PTFE clamps and are thermally coupled to a heat bath at $\sim$10~mK by means of a copper holder structure.
To enhance the light collection, the lateral sides of ZnSe crystals have been surrounded with reflecting foils, later removed in the second phase of data taking.
The reader can find some pictures of the detector in Fig.~\ref{Fig:Detector} and more details in Ref.~\cite{CUPID0:Detector:2018tum}.

\begin{figure}[t!]
\begin{center}
\subfigure{\includegraphics[height=0.3\textwidth]{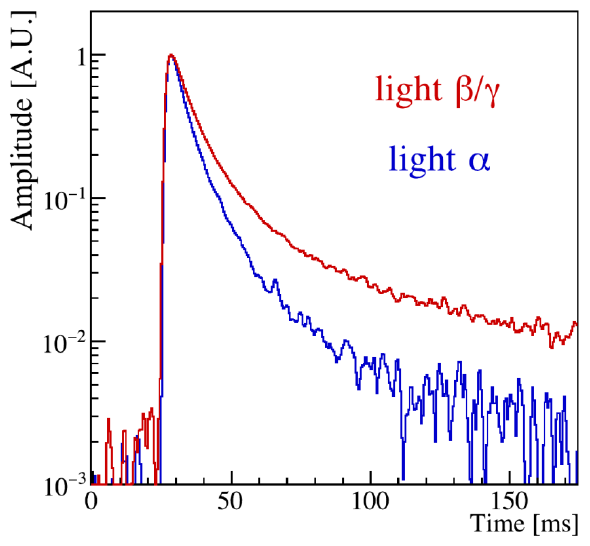}}
\hspace{2cm}
\subfigure{\includegraphics[height=0.3\textwidth]{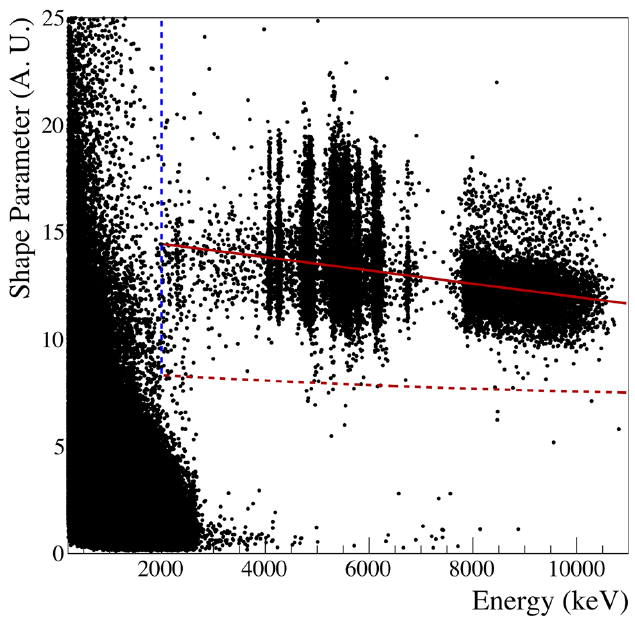}}
\end{center}
\caption{Left: shapes of the light pulses produced by $\alpha$ and $\beta/\gamma$ particle interactions. Right: scatter plot of the shape parameter as a function of energy, exploited to discriminate $\alpha$ particles at E$>$2~MeV.}
\label{Fig:ParticleID} 
\end{figure}

In the first scientific run, that lasted from March 2017 until December 2018, CUPID-0 collected an exposure of 9.95~kg$\times$yr with an excellent duty cycle (the time fraction devoted to $\beta\beta$ physics and calibrations is 88\%).
$^{232}$Th sources were periodically deployed besides the cryostat to perform the energy calibration of the heat pulses and the inter-calibration of light detectors. 
Moreover a calibration with a $^{56}$Co source was performed to check the goodness of energy reconstruction and evaluate the resolution at $^{82}$Se Q$_{\beta\beta}$, which resulted equal to 20.05$\pm$0.34~keV after removing the correlation between
light and heat signals~\cite{Beretta:2019bmm}.
The physics spectrum is built after applying a series of selection criteria aimed at improving the experimental sensitivity~\cite{CUPID0:Analysis:2018yye}. 
First, we reject pile-up events (1~s before and 4~s after trigger) and we select only signals consistent with a proper template waveform in order to identify real particle events. 
Then, we tag the events that simultaneously trigger more than one crystal within a $\pm$20~ms time window, assigning a \textit{multiplicity} label equal to the number of crystals hit. Indeed, according to our Monte Carlo simulations, a potential 0$\nu$DBD event is expected to release all its energy in a single crystal with a probability of 81\%, thus the modularity of the detector allows to reduce the background due to particle interactions depositing energy in multiple crystals. 
Finally, we perform the particle identification by analyzing the time-development of light pulses. As shown in Fig.~\ref{Fig:ParticleID}, the shape of light pulses produced by an $\alpha$ interaction is different from that produced by $\beta/\gamma$ particles. By exploiting this feature, we separate $>$99.9\% of $\alpha$ events from $\beta/\gamma$ ones at energies $>$2~MeV. 

\section{Background analysis and physics results}

To measure the background around the $^{82}$Se Q$_{\beta\beta}$ with significant statistics, we select a wide energy region from 2.8 to 3.2~MeV. 
By applying $\alpha$ discrimination, the background level decreases from (3.2$\pm$0.3)$\times10^{-2}$ to (1.3$\pm$0.2)$\times10^{-2}$ counts/(keV~kg~yr). 
To further suppress the background in this region, we tag potential $^{212}$Bi $\alpha$ decays and we veto any event occurring within 7 half-lives of its daughter $^{208}$Tl ($T_{1/2}$=3.05~min) that $\beta$ decays with a high Q-value (5~MeV). This technique allows to reduce the background down to $(3.5^{+1.0}_{-0.9})\times10^{-3}$ counts/(keV~kg~yr), at the cost of only 6\% dead time.
Since we find no evidence for $^{82}$Se 0$\nu$DBD, we perform an Unbinned Extended Maximum Likelihood fit in the [2.8$-$3.2]~MeV range to set the most stringent lower limit on the half-life of this process: $T^{0\nu}_{1/2}>3.5\times10^{24}$~yr (90\% credible interval) \cite{CUPID0:0nu:2019tta}. 

\begin{figure}[t!]
\begin{center}
\subfigure{\includegraphics[height=0.3\textwidth]{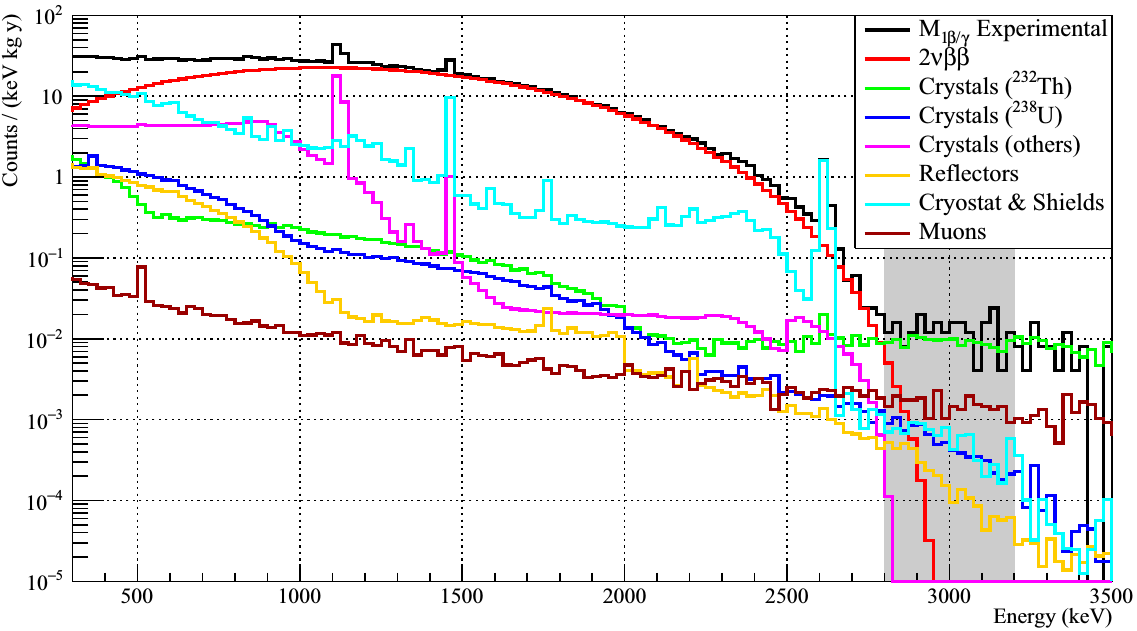}}
\hspace{0.5cm}
\subfigure{\includegraphics[height=0.3\textwidth]{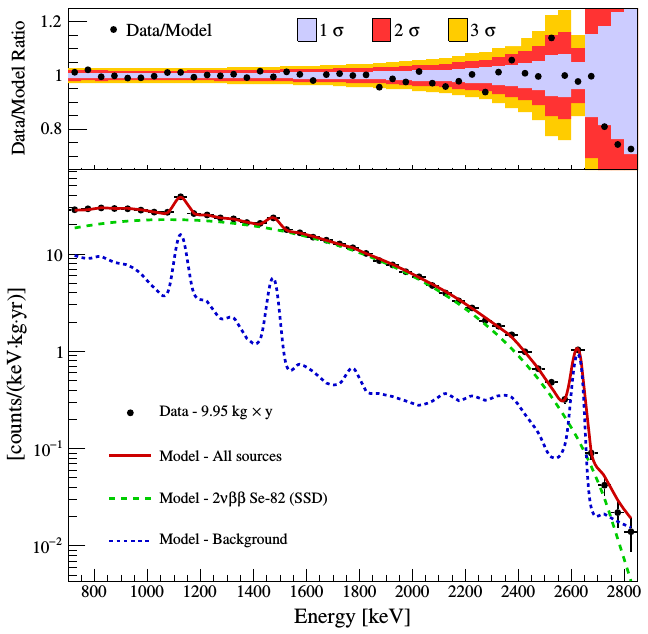}}
\end{center}
\caption{Left: background sources contributing to the spectrum of $\beta/\gamma$ events hitting one single crystal. The shaded area corresponds to the energy range from 2.8 to 3.2 MeV chosen to analyze the background around the $^{82}$Se Q$_{\beta\beta}$. In this plot, the time veto for the rejection of $^{208}$Tl events is not applied. Right: BM fit with the SSD hypothesis for the 2$\nu$DBD. In the top panel, we show the bin-by-bin ratio between experimental and reconstructed counts with the uncertainties at 1, 2, 3$\sigma$ shown as colored bands centered at 1.}
\label{Fig:BM2nu} 
\end{figure}

In order to understand the origin of the background in CUPID-0 and analyze the 2$\nu$DBD signal, we create a model based on a tool developed in the framework of the CUORE-0 experiment~\cite{CUORE-0:2016vtd, CUPID0:BM:2019nmi}. 
We divide the data according to their multiplicity and particle type and we analyze the $\gamma$ and $\alpha$ lines in the spectra to recognize the signatures of the background sources, i.e.~contaminations of the experimental setup by natural or anthropogenic radioisotopes. 
Then, we run Geant4-based Monte Carlo (MC) simulations of the background sources to get the energy spectra produced in the detector by each of them. 
Finally, we perform a Bayesian fit that stacks the MC spectra with free normalization coefficients to reproduce the experimental data (Fig.~\ref{Fig:BM2nu}a).
Through this analysis we obtain that $\sim44\%$ of the residual background around $^{82}$Se Q$_{\beta\beta}$ is produced by cosmic muon showers, while the remaining fraction is due to radioactive contaminations of ZnSe crystals ($\sim33\%$), of cryogenic setup ($\sim17\%$), and of reflector foils and copper holder ($\sim6\%$).

The CUPID-0 background model (BM) has been exploited also to measure the 2$\nu$DBD of $^{82}$Se with unprecedented precision and accuracy: $T^{2\nu}_{1/2} = [8.60 \pm 0.03 (\text{stat}) ^{+0.19}_{-0.13} (\text{stat})]\times10^{19}$~yr~\cite{CUPID0:2nu:2019yib}. 
In this work, we test two different nuclear models for the 2$\nu$DBD, which can be described as a sequence of two virtual $\beta$ decays going through one or more states of the (A,Z+1) intermediate nucleus. 
Depending on the model, the 2$\nu$DBD is referred to as \textit{Single$-$State Dominated} (SSD) or \textit{Higher$-$States Dominated} (HSD) and its energy spectrum turns out to be slightly different. 
Thanks to the excellent signal to background ratio in CUPID-0 data (Fig.~\ref{Fig:BM2nu}b), we have a strong evidence that the 2$\nu$DBD of $^{82}$Se is SSD, because the counts predicted by the BM fit when using the HSD hypothesis is not compatible at $5.5\sigma$ with the experimental data in the 2$-$3~MeV range.
Furthermore, through another analysis of the CUPID-0 2$\nu$DBD spectral shape, we could set a limit on CPT violation in the $\beta\beta$ decay of $^{82}$Se~\cite{CUPID0:CPTV:2019swx}.

In June 2019, we started the second scientific run of CUPID-0, accumulating 5.2 kg$\times$yr exposure, with the aim of better identifying background contributions in view of CUPID. For this purpose, we installed a muon veto and we removed the reflecting foils around crystals in order to enhance the capability of measuring their surface contaminations.

\section{Conclusion}
The CUPID-0 experiment demonstrated that the dual readout of heat and light allows to reach the lowest background for cryogenic calorimeters, laying a solid foundation for CUPID. 

Despite the limited mass, CUPID-0 was able to establish the best half-life limit on $^{82}$Se 0$\nu$DBD and the most precise measurement of $^{82}$Se 2$\nu$DBD half-life, unveiling that the nuclear mechanism mediating this process is SSD.

\begin{footnotesize}
\section*{\small Acknowledgements} 
This work was partially supported by the Low-background Underground Cryogenic Installation For Elusive Rates (LUCIFER) experiment, funded by ERC under the European Union's Seventh Framework Programme (FP7/2007-2013)/ERC grant agreement n. 247115, funded within the ASPERA 2nd Common Call for R\&D Activities, and was funded by the Istituto Nazionale di Fisica Nucleare. We thank M.~Iannone for his help in all the stages of the detector assembly,  A.~Pelosi for constructing the assembly line, M. Guetti for the assistance in the cryogenic operations, R. Gaigher for the mechanics of the calibration system, M. Lindozzi for the cryostat monitoring system, M. Perego for his invaluable help in many tasks, the mechanical workshop of LNGS (E. Tatananni, A. Rotilio, A. Corsi, and B. Romualdi) for the continuous help in the overall set-up design. A.~S.~Zolotorova is supported by the Initiative Doctorale Interdisciplinaire 2015 project funded by the Initiatives d'excellence Paris-Saclay, ANR-11-IDEX-0003-0. We acknowledge the Dark Side Collaboration for the use of the low-radon clean room. This work makes use of the DIANA data analysis and APOLLO data acquisition software which has been developed by the CUORICINO, CUORE, LUCIFER and CUPID-0 collaborations.\par
\end{footnotesize}

\bibliographystyle{JHEP}
\bibliography{Bibliography}

\end{document}